# Exploring the Cognitive Dynamics of Artificial Intelligence in the Post-COVID-19 and Learning 3.0 Era: A Case Study of ChatGPT

Lingfei Luan[1], Xi Lin[2],Wenbiao Li[1]

[1]Case Western Reserve University
[2]East Carolina University

February 3rd, 2023

## Abstract
The emergence of artificial intelligence has incited a paradigm shift across the spectrum of human endeavors, with ChatGPT serving as a catalyst for the transformation of various established domains, including but not limited to education, journalism, security, and ethics. In the post-pandemic era, the widespread adoption of remote work has prompted the educational sector to reassess conventional pedagogical methods. This paper is to scrutinize the underlying psychological principles of ChatGPT, delve into the factors that captivate user attention, and implicate its ramifications on the future of learning. The ultimate objective of this study is to instigate a scholarly discourse on the interplay between technological advancements in education and the evolution of human learning patterns, raising the question of whether technology is driving human evolution or vice versa.
   *Keywords*：Artificial intelligence (AI), Human-machine communication, COVID-19, Chat GPT, Learning 3.0, Critical Thinking

## 1.Introduction of ChatGPT
ChatGPT, a chatbot developed by OpenAI, can interpret and respond to natural language input using the GPT-3 language model which has 175 billion parameters (Floridi & Chiriatti, 2020). The utilization of a word-driven dialogue system offers assistance in cross-domain problem resolution and the generation of content to answer users' inquiries. (old: Word-driven dialogue provides support with cross-domain problem-solving and generative content to address users' questions). Within one week of its introduction to the public, one million consumers had signed up for the platform (Haque et al., 2022). This has spurred a discussion over the implications of ChatGPT in a variety of fields, including education. Some educators claim that the emergence of such technology renders conventional online exams outmoded and raises worries over the eventual automation of the teaching profession (Kung et al., 2022; Cotton, 2023; Pavlik, 2023; King &ChatGPT 2023). In contrast, some argue that ChatGPT has the potential to improve students' adaptability to changing educational needs and develop independent learning practices, which also assist instructors in providing customized training plans (Zhai, 2022; Qadir, 2022; FIRAT, 2022). The considerable interest received by this chatbot necessitates more research into the psychological underpinning its use and its possible influence on the evolution of educational paradigms. The discourse around AI's incorporation into the education sector and its role in molding the education of future generations include not just talks of technology developments, but also considerations of the human experience and the behavioral consequences of technological change.

## 2.1. ChatGPT as the new force to reshape humans' preferences in social needs

The Computers Are Social Actors (CASA) paradigm claims that the relationship between people and computers is one of social interaction rather than tool use. This concept suggests that computers, similar to human social actors, possess the power to influence and modify human behavior and social interactions (Reeves and Nass, 1996). Contrary to the conventional view of computers as passive tools, this approach views computers as active participants in human relationships. Lee et al. (2006) indicate that people subconsciously adopt the behavioral standards governing human relationships in their interactions with machines. Their argument implies that people consider computers as social actors and respond to them similarly to how they would behave to other human actors. Forsyth (2010) extends this perspective by pointing out that interpersonal interaction can serve as either task- or social-oriented communication goals. The social-oriented goal focuses on sustaining social connections and interactions, whereas task-oriented behavior focuses on accomplishing a specific task in an optimized approach. Interpersonal interaction may approach human-machine interaction differently depending on the communication objectivity and orientation, for example, some may emphasize connection orientation and focus on human self-esteem and well-being, whilst others may favor functional orientation and optimize the action plan to complete a specific task.

In effect, Forsyth's argument implies that interpersonal exchange contains a distinguishing two systems including fundamental goals, relationship cohesion, efficacy, relationship satisfaction, and even other implements such as patient care provided in healthcare settings (Myers et al., 1999; Tabernero et al., 2009; Bruno et al., 2017; Rüzgar, 2018). This shows that the method by which humans interact with computers may vary dependent on context or relationship type and that it is essential to investigate human-computer interactions by examining multiple systems. With the introduction of ChatGPT, this differentiation between the two independent systems has been totally eliminated. ChatGPT is a generative AI solution with both functional and social features, satisfying the demands of individuals who wish to get answers fast and functioning as an optimal assistant to provide personalized and interactive support.

First, ChatGPT preserves the AI-powered chatbot design in terms of functionality but creates appropriate answer patterns depending on user input, so making online text communication more natural and intuitive (OpenAI, 2023). While the controversy persists over whether utilizing ChatGPT to write an essay or assignment accounts for plagiarism, it cannot be denied that ChatGPT utilizes the strength of the natural language training model to satisfy the demands of most domains. Students may use ChatGPT for multiple purposes, for instance, to request information towards certain topics, summarize articles, provide writing revisions, or establish a logical core connection or outline for a paper via multiple questions. By delivering individualized and interactive assistance to meet the individual requirements and preferences of each user, this model represents the epitome of generative AI. Real-time feedback and personalized assistance can empower users in making more rapid and effective adjustments, hence facilitating job completion and performance. Second, the integration characteristics of ChatGPT offer users a more flexible and simple search channel comparing traditional searching. Users do not need to go through several large databases or platforms to locate relevant information. Instead, they simply provide ChatGPT with precise directions during conversations, such as looking for measurements of protein molecules in academic databases and listing the relevant articles. This strategy enhances the utilization of open educational resources (FIRAT, 2022). According to the learning materials and resources to give individualized guidance, ChatGPT can comprehensively satisfy the demands of users, and users to seek the answer to deliver the optimal solution. Based on customized guidance supplied by learning materials and resources, ChatGPT can not only fulfill the demands of users in all aspects but also provide the best solution for users seeking answers. As a result, ChatGPT has the most advanced functional characteristics in human-computer interaction. Secondly, ChatGPT redefines the role and function of social robots in human-social interaction, addressing people's emotional requirements to develop a feeling of autonomy and anti-anxiety in the post-COVID-19 period. In accordance with Forsyth's concept of distinct human-machine systems, artificial intelligence has historically developed two distinct products: functional AI, which aims to enhance efficiency in task performance, such as domestic chores (Reiser et al., 2013), and companion robots, which fulfill emotional needs for companionship, such as Nanny robots. Social robots, on the other hand, are designed to promote humans' well-being and provide companionship (Kanamori et al., 2002). Individuals have a more positive attitude toward functional robots built for home duties since they have the ability to boost workers' productivity, as evidenced by empirical study (Dautenhahn, 2005; Ray et al., 2008; Kim et al., 2021). Kim et al. (2021) conducted a study to explore humans' attitudes toward two types of AIs (Functional and Social). Participants were given two three-minute video clips, one labeled "functional" and the other "social", introducing an AI character Samantha. The purpose of the study was to assess the views of participants about AI, with a focus on the differences between those who viewed the "functional" video and those who viewed the "social" film. Compared to the "social" group, participants in the "functional" group had a more positive attitude toward AI, as demonstrated by the study's findings. It is important to note, however, that the study did not seek to determine whether the participants had actually experienced or interacted with different types of AI in the past; consequently, the results can only be interpreted as a preference for the content of the two AI videos, rather than an actual experience with AI. In addition, the study's analysis of the experiment is confined to the preference of the two groups of subjects for the content of the two AI movies and cannot provide insight into the participants' real experience with AI.

Based on the Relationships Motivation Theory, autonomy and motivation are essential aspects of human development and advancement (Deci et al., 2014). Specifically, Individuals seek autonomy in order to satisfy their physiological, security, emotional, dignity, and self-actualization needs (Maslow, 1943). As an artificial intelligence-based system, ChatGPT has the potential to aid in the satisfaction of emotional, dignity, and self-actualization requirements. This is achieved partly through ChatGPT's 24-hour and instant service, which enables individualized user interactions through ChatGPT's 24-hour and instant service, which enables individualized interactions with users. Answering users' inquiries individually can promote a feeling of social camaraderie and care. In addition, the content of these interactions remains private, allowing users to feel appreciated and valued without fear of the conversation being criticized or exposed. In addition, the customized content of ChatGPT can assist users in swiftly locating answers to problems, therefore enhancing their productivity and performance and contributing to a greater sense of dignity. Self-actualization being the highest degree of human need, ChatGPT is able to give a vast array of intelligent assistance to make individuals feel more productive and purposeful in their job, hence fostering feelings of self-worth. For example, students may receive good grades for writing papers brainstorming with and being guided by ChatGPT, and programmers can receive rewards for optimizing code using ChatGPT.

## 2.2 The influence of COVID-19 on online education

The shift towards remote instruction in higher education institutions, as a result of the COVID-19 pandemic, has led to a significant increase in the use of online learning and the number of students enrolled in online courses. Allen and Seaman (2010, 2013) reported a consecutive seven-year growth in student online enrollment, with a 21% growth rate, and a 32% participation rate of college students in at least one online course. Despite the potential benefits of online learning, such as increased autonomy and flexibility, students often face various challenges, particularly in terms of self-regulation and independence when studying online (Cho et al., 2010). Self-regulated learning is a concept that refers to the self-directed processes and self-perceptions that enable learners to effectively utilize their cognitive abilities and achieve academic success (Zimmerman, 2008). The theory of self-regulated learning, developed by Albert Bandura in his social-cognitive learning theory, posits that human functioning is a dynamic interaction between personal, behavioral, and environmental factors, referred to as triadic reciprocity (Bandura, 1986). Specifically, personal factors include individual self-efficacy, goal orientation, and metacognition, environmental factors include instruction, peer learning, and help-seeking in an online learning context, and behavioral factors focus on learning performance (Schraw et al., 2006). Researchers have emphasized that self-regulated learning, as an active and constructive process, involves goal orientation, self-efficacy, self-control, motivation, cognitive strategies, and metacognitive self-regulation (Pintrich & Zusho, 2002). As a result, self-regulated learners tend to be self-motivated, use metacognitive learning strategies frequently, and exhibit high academic performance (Zimmerman & Martinez-Pons, 1986).

Additionally, the concept of self-regulated learning involves the ongoing adjustment of cognitive processes and activities to align with the specific learning context (Garcia & Pintrich, 1991). Zimmerman and Schunk (2001) further noted that self-regulation is highly dependent on the context of learning. In the case of online learning, the lack of real-time interaction with instructors and physical distance from school resources can create academic and emotional challenges for students (Bowers & Kumar, 2015). However, online learning environments offer students greater autonomy, and it is expected that they will be independent and self-regulated in order to maintain cognitive engagement and motivation (McMahon & Oliver, 2001). To address these challenges, virtual mentoring utilizing AI has been suggested as a solution to support student learning (Carter et al., 2020). A study by Siemens (2013) found that AI coaching tools have the potential to help learners effectively use self-regulated learning strategies and detect when they need assistance. Additionally, AI tutoring systems have been in use since 1997 and have played a crucial role in online learning. Hwang and colleagues' review of AI-supported online learning stated that the combination of intelligent tutoring systems and distance learning began in 1997, and with the advancement of AI algorithms, AI tutoring systems have improved learner-system interactions to promote personalized, adaptive, and collaborative learning.

**2.3 ChatGPT in Learning 3.0 Era: Definition, Implications, and the Debate on Technological vs Human Evolution**

The term "Learning 3.0" was introduced by Maria Langworthy (Langworthy & Hirsch-Allen, 2022). It describes a new trend in learning and education, which encompasses a transition from traditional teacher-centered learning to more student-centered with the assistance of technology. Learning 3.0 emphasizes critical thinking, creativity, and problem-solving, while Learning 1.0 emphasizes memorization and the acquisition of facts and information, and Learning 2.0 is a transition to a technology-enabled and student-centered approach that focuses on knowledge application and skills like collaboration, communication, and problem-solving. The core elements of Learning 3.0 include skills-based education and verification for career mobility, data-driven personalized education for career development, decentralized education with micro-learning and quality assessment, and revenue from the skills-focused education model (Langworthy & Hirsch-Allen, 2022). From the core elements of Learning 3.0, it is clear that the education system is adapting to the needs of social development, as the emphasis has shifted from obtaining degrees to building human and technological systems and catering to the requirements of a rapidly evolving work and life. Creativity, communication, adaptability, self-awareness, autonomy, critical thinking, and teamwork are at the heart of Learning 3.0's emphasis on long-lasting abilities for the next generation. Namely, the objective of Learning 3.0 is to create a customized and flexible learning and working system, which may be everlasting (Roslansky, 2021). Is ChatGPT a component of this trend or Learning 3.0? Have people actively evolved the educational system, or has the system changed with new technology over time?

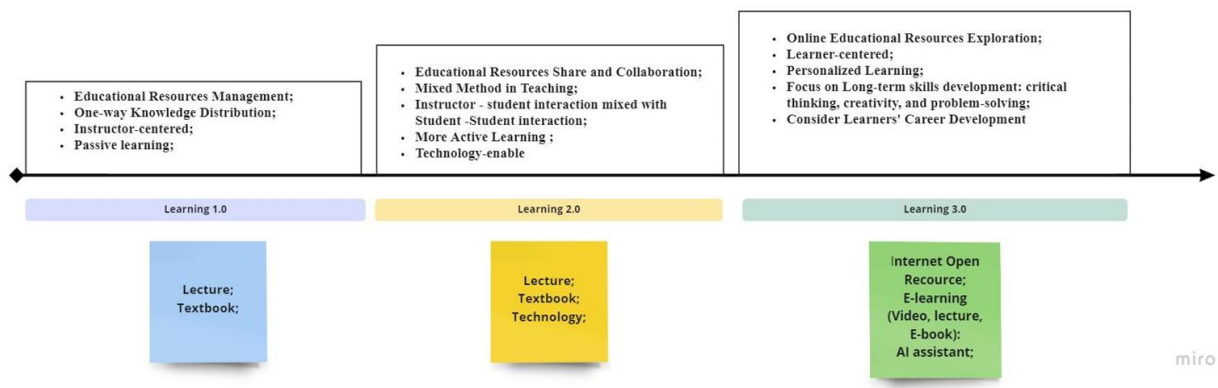

Figure 1. The Development of Learning and Their Core Components

With the expansion of Internet technology, particularly after Covid-19, online education has flourished, and Internet technology has grown inseparable from learners. Both students and instructors are faced with a daunting task: interacting with the Internet's massive volume of information and selecting the finest information. These obstacles impact not just teaching methods but also learning methods and the effectiveness of learning. When students can discover answers to simple queries in a few seconds through ChatGPT, the conventional one-way approach for disseminating educational information is outdated. The Learning 3.0 abilities of adaptability, autonomy, collaboration, and communication are expressed in the Q & A between users and ChatGPT. Similarly, the core concept of Learning 3.0 is to be student-centered and to deliver individualized educational objectives. When users are interacting with ChatGPT, users are the subjects of the interaction for learning, information acquisition, and problem-solving. The solution-driven interaction actually promotes the users' autonomy of learning and enhances their adaptability to collaborative learning as well as encourages students to seek the best answers to their questions and delve deeply into their knowledge. AI now embodies the dual attributes of friend and helper; it not only demonstrates its capability as a natural language-based information exchange and generating service for users, but also as a 24-hour partner willing to answer any queries at any time. This may be the primary reason why ChatGPT obtained a huge number of users in a short amount of time, breaching the constraints of traditional AI design and presenting itself to the market as a convergent product.

If we consider the transmission and exchange of educational resources as the construction and operation of an information system, then Learning 3.0's innovation is to switch the traditional one-way information dissemination (teachers-centered) to comprehensive knowledge discovery and interaction (student-centered). The induction of ChatGPT provides the approaches and ability for users to navigate through the transition. According to Davis' technology acceptance model (TAM), another reason for ChatGPT's quick adoption is that it offers two criteria for tool awareness: perceived usefulness (PU) and perceived ease of use (PEU) (Davis, 1985). There is no doubt that ChatGPT users will increase their job and study efficiency, save their time, and improve their performance and quality, all of which are emphasized by TAM as advantageous. With ChatGPT's straightforward question-and-answer structure, users do not need to acquire particular skills to obtain the necessary information. This energy-saving consumption strategy reflects TAM's user-friendliness. Yang and Yoo (2004) found that cognitive attitudes alter the influence of PU and PEU on information system usage. In this case, the attitudes of users and educators toward new technologies and innovations will have a significant impact on the promotion of Learning 3.0 and the role of ChatGPT inside the educational system. In conclusion, it is not so much that the educational system changes with the times as that new technology and the enhancement of people's cognitive awareness produce new instruments that encourage the evolution of the educational system.

**Conclusion**
The outbreak of COVID-19 has had a significant influence on human development, particularly education. In the post-epidemic age, both instructors and students face uncertainties in the teaching and learning process along with employment pressure. Under the combined constraints of education and employment, educators and students must reevaluate current educational approaches and outcomes, as well as the future career development of students. As a companion and task-accomplishing AI, ChatGPT opens up a new world of learning for users by providing them with rapid, efficient, and individualized assistance and support. From Q&A with ChatGPT, one can easily obtain an appropriate solution or crucial information, and thereby gain the motivation to eliminate anxiety caused by uncertainty in learning, while enhancing interest and self-exploration in order to achieve psychological satisfaction and peace. Furthermore, this customized exploration enables students, who are self-directed and focused on their own future development, to gain access to more useful learning resources. In general, ChatGPT expands the options available to all users, whether it is through the introduction of new technologies or changes to the educational system. Individuals are addicted to continually investigating new uses of ChatGPT and implementing them in practice, such as resume writing and cover letter evaluation. In sum, this study aims to share further explanations and theoretical foundations for future research by investigating the psychological process underlying ChatGPT, as well as enlightening educators on how to properly view AI and ChatGPT.